\documentstyle[12pt]{article}

\title{
{\vspace{-3cm} \normalsize
\hfill \parbox{30mm}{DESY 93-036 \\
                     MS-TPI-93-01 \\
                     SHEP-92/93-15}  }\\[25mm]
Bounds on the renormalized couplings in an
$\rm SU(2)_L \otimes SU(2)_R$ symmetric Yukawa model}

\author{
 {L. Lin}\thanks{Institut f\"ur Theoretische Physik I,
  Universit\"at M\"unster, Wilhelm-Klemm-Str.~9, D-4400 M\"unster, FRG}
 \and
 {I. Montvay}\thanks{Deutsches Elektronen-Synchrotron DESY,
                 Notkestr.\,85, D-2000 Hamburg 52, FRG}
\and
 \addtocounter{footnote}{-1}%
 {$\mbox{G. M\"unster}^{\scriptstyle
 \fnsymbol{footnote} }$}
\and
 {$\mbox{M. Plagge}^{\scriptstyle
 \fnsymbol{footnote} }$}
\and
 \addtocounter{footnote}{+1}%
 {H. Wittig}\thanks{Department of Theoretical Physics,
                    University of Southampton, UK}
 }

\date{March 23, 1993}

\newcommand{\be}{\begin{equation}}
\newcommand{\ee}{\end{equation}}
\newcommand{\half}{\frac{1}{2}}

\begin{document}
\maketitle

\begin{abstract} \normalsize
The vacuum stability lower bound on the mass of the Higgs boson
is numerically investigated in an $\rm SU(2)_L \otimes SU(2)_R$
symmetric Yukawa model, which describes two heavy degenerate fermion
doublets in the limit of vanishing gauge couplings.
Good agreement with perturbation theory is found, although the
couplings are strong.
The upper bound on the fermion mass and renormalized Yukawa coupling
is also determined in the part of bare parameter space where
reflection positivity has been proven.
\end{abstract}
\vspace{1cm}


\section{Introduction}                                    \label{s1}

Cut-off dependent upper bounds on renormalized couplings arise in the
Standard Model for non-asymptotically free couplings, if their
continuum limit is trivial.
Such ``triviality bounds'' show up in perturbation theory as an
apparent inconsistency \cite{LANPOM}, and have been intensively studied
on the lattice by non-perturbative methods (for a recent review see
\cite{PETCH}).
In Yukawa models there is also a lower limit on the Higgs mass,
which is called ``vacuum stability bound'' \cite{DUPHSH}.
On the lattice this can be understood as due to quantum effects
inducing a positive renormalized quartic scalar coupling even if
the bare quartic coupling takes its lowest possible value, namely
zero \cite{LMMW}.

In previous papers \cite{FLMMPTW} the triviality upper bound on the
Higgs mass has been investigated by non-perturbative methods in a
Yukawa model with chiral $\rm SU(2)_L \otimes SU(2)_R$ symmetry.
The lattice formulation is based on the mirror fermion action
\cite{CHFER} with exact decoupling of the mirror fermions from the
physical spectrum \cite{ROMA1,GOLPET}.
(Another lattice formulation of the same continuum ``target theory''
is possible by using reduced staggered fermions \cite{BOSMVI}.)
Since the Hybrid Monte Carlo method \cite{DKPR} is used, the minimum
number of fermion doublets which can be numerically simulated, is two.
In this context it is important that we are working in the limit of
vanishing gauge couplings, and hence by charge conjugation
a mirror fermion doublet can be transformed to a fermion doublet.
In fact, only the vanishing of the
$\rm SU(3)_{colour} \otimes U(1)_{hypercharge}$ gauge couplings is
necessary.
The chiral $\rm SU(2)_L$ gauge coupling can be introduced, because
SU(2) is pseudoreal \cite{ROMAPR,AMSTER}.

In the present paper we continue the non-perturbative investigation
of the same $\rm SU(2)_L \otimes SU(2)_R$ symmetric Yukawa model
as in refs.\ \cite{LINWIT,FLMMPTW}.
In particular, we concentrate on the vacuum stability lower bound
and on the upper limit of the renormalized Yukawa coupling in the
region of positive scalar hopping parameter, where reflection
positivity, implying unitarity in Minkowski space, can been proven
\cite{LMMW}.
Besides the physically relevant phase with broken symmetry, the
renormalized Yukawa coupling is also computed in the symmetric phase,
in order to check previous observations of very strong couplings
there \cite{FKLMMM,LINWIT}.


\section{Vacuum stability bound}                          \label{s2}

The lattice action and the definition of different renormalized
physical quantities closely follow our previous papers on chiral
Yukawa models with mirror fermions \cite{FKLMMM,LMMW,LINWIT,FLMMPTW}.
The lattice action is a sum of the O(4)
($\cong \rm SU(2)_L \otimes SU(2)_R$) symmetric pure scalar part
$S_\varphi$ and fermionic part $S_\Psi$:
\be \label{eq01}
S = S_\varphi + S_\Psi \,.
\ee
$\varphi_x$ is the $2 \otimes 2$ matrix scalar field, and
$\Psi_x \equiv (\psi_x, \chi_x)$ stands for the mirror pair of fermion
doublet fields (usually $\psi$ is the fermion doublet and $\chi$ the
mirror fermion doublet).
In the usual normalization conventions for numerical simulations we
have
$$
S_\varphi = \sum_x \left\{ \half {\rm Tr\,}(\varphi_x^+\varphi_x) +
\lambda \left[ \half{\rm Tr\,}(\varphi_x^+\varphi_x) - 1\right]^2
- \kappa\sum_{\mu=1}^4
{\rm Tr\,}(\varphi^+_{x+\hat{\mu}}\varphi_x)
\right\} \ ,
$$
$$
S_\Psi = \sum_x \left\{ \mu_{\psi\chi} \left[
(\overline{\chi}_x\psi_x) + (\overline{\psi}_x\chi_x) \right]
\right.
$$
$$
- K \sum_{\mu=\pm 1}^{\pm 4} \left[
(\overline{\psi}_{x+\hat{\mu}} \gamma_\mu \psi_x) +
(\overline{\chi}_{x+\hat{\mu}} \gamma_\mu \chi_x)
+ r \left( (\overline{\chi}_{x+\hat{\mu}}\psi_x)
- (\overline{\chi}_x\psi_x)
+ (\overline{\psi}_{x+\hat{\mu}}\chi_x)
- (\overline{\psi}_x\chi_x)  \right) \right]
$$
\be \label{eq02}
\left.
+ G_\psi \left[ (\overline{\psi}_{Rx}\varphi^+_x\psi_{Lx}) +
(\overline{\psi}_{Lx}\varphi_x\psi_{Rx}) \right]
+ G_\chi \left[ (\overline{\chi}_{Rx}\varphi_x\chi_{Lx}) +
(\overline{\chi}_{Lx}\varphi^+_x\chi_{Rx}) \right]
\right\} \,.
\ee
Here $K$ is the fermion hopping parameter, $r$ the Wilson-parameter,
which will be fixed to $r=1$ in the numerical simulations, and the
indices $L,R$ denote, as usual, the chiral components of fermion
fields.
In this normalization the fermion--mirror-fermion mixing mass is
$\mu_{\psi\chi}=1-8rK$.

The fermionic part $S_\Psi$ is given here for a single mirror
pair of fermions.
For the Hybrid Monte Carlo simulation the fermions have to be
doubled by taking the adjoint of the fermion matrix for the new
species.
Taking the adjoint transforms fermions to mirror fermions and vice
versa, but as noted before, without
$\rm SU(3)_{colour} \otimes U(1)_{hypercharge}$ gauge couplings they
are equivalent to each other.

The numerical simulations were performed on $6^3 \cdot 12$
lattices at a bare scalar quartic coupling $\lambda=10^{-6}$.
The small positive value of $\lambda$ was chosen in order to be
sure about the convergence of the path integral.
The Yukawa coupling $G_\chi$ was kept at zero, for exact decoupling of
the mirror doublets \cite{GOLPET}.
This allows to stay with the fermion hopping parameter $K$ near its
critical value at $K=1/8$, as described in ref. \cite{FLMMPTW}.
In the broken (FM) phase at the fixed values $G_\psi=0.25$ and
$G_\psi=0.30$ the scalar hopping parameter was tuned to achieve a
scalar mass of about $m_{R\sigma} \simeq 0.6 - 0.8$ and a not too small
fermion mass.
$\kappa$ was always kept to be non-negative, in order to be sure
about reflection positivity, that is unitarity in Minkowski
space \cite{LMMW}.
The last points in the positivity region were fixed at $\kappa=0$,
and then $G_\psi$ was tuned to obtain reasonable masses.
The results are summarized in table \ref{t1}, where also such points
are included where the masses are not yet sufficiently tuned.

%
\begin{table}[tb]
%
%
\caption{  \label{t1}
The main renormalized quantities and the bare magnetization
$\langle \sigma \rangle \equiv \langle|\varphi|\rangle$ for several
bare couplings $G_\psi$ and $\kappa$-values.
Points labelled by small letters are our data at $\lambda=10^{-6}$ which
were all obtained on $6^3\cdot12$.
Points labelled by capital letters are the data for the upper bound at
$\lambda=\infty$.
For the points with double capital letters the lattice size was
$8^3\cdot16$, whereas single capital letters denote data on
$6^3\cdot12$.
All data are collected from typically 10000 trajectories, except for the
points labelled a, i, A, C where only about 5000 trajectories were run.
As $G_\chi=0$, the renormalized coupling $G_{R\chi}$ was always zero
within errors and is not included here.}
\begin{center}
\begin{tabular}
{|c|r@{.}l|r@{.}l|r@{.}l|r@{.}l|r@{.}l|r@{.}l|c|r@{.}l|r@{.}l|}
\hline
&\multicolumn{2}{c|}{$G_\psi$} & \multicolumn{2}{c|}{$\kappa$}
&\multicolumn{2}{c|}{$\langle\sigma\rangle$}
&\multicolumn{2}{c|}{$v_R$}
&\multicolumn{2}{c|}{$m_{R\sigma}$} &\multicolumn{2}{c|}{$\mu_{R\psi}$}
& $g_R$ & \multicolumn{2}{c|}{$G_{R\psi}$}
& \multicolumn{2}{c|}{$G_{R\psi}^{(3)}$} \\
\hline
 a & 0&25 &  0&095 & 0&251(7) & 0&196(11)  & 0&79(7)
   & 0&271(8)& 42(11)  &  1&38(4) &  1&42(20) \\
 b & 0&25 &  0&099 & 0&393(7) & 0&241(10)  & 0&59(9)
   & 0&407(9)& 18(3)   &  1&69(4) &  1&56(16) \\
 c & 0&25 &  0&101 & 0&525(6) & 0&304(9)   & 0&57(3)
   & 0&59(3) & 10(2)   &  1&68(10) &  1&53(40) \\
\hline
 d & 0&3  &  0&090 & 0&438(6) & 0&27(1)    & 0&75(7)
   & 0&55(6) & 23(3)   &  2&04(10) & \multicolumn{2}{c|}{}  \\

 e & 0&3  &  0&095 & 0&754(4) & 0&417(11)  & 0&76(7)
   & 0&91(6) & 10(1)   &  2&19(11) & \multicolumn{2}{c|}{}  \\
 f & 0&3  &  0&100 & 1&260(5) & 0&67(2)    & 0&84(6)
   & 1&503(12)& 4.8(8)  &  2&24(7) & \multicolumn{2}{c|}{}  \\
\hline
 g & 0&62 &  0&0   & 0&424(3) & 0&29(2)    & 1&23(9)
   & 1&23(3) & 53(11)  &  4&2(1.2) &  4&5(4.0)  \\
 h & 0&63 &  0&0   & 0&469(3) & 0&351(11)  & 1&65(13)
   & 1&25(16)& 63(16)  &  3&5(4) & \multicolumn{2}{c|}{}  \\
 i & 0&65 &  0&0   & 0&523(3) & 0&34(2)    & 1&39(14)
   & 1&6(3)  & 53(18)  &  4&6(4) & \multicolumn{2}{c|}{}  \\
\hline
\hline
 A & 0&3  &  0&30  & 0&439(1) & 0&400(13)  & 1&17(7)
   & 0&55(2) & 26(7)   &  1&36(6) & \multicolumn{2}{c|}{}  \\
BB & 0&3  &  0&27  & 0&270(2) & 0&25(1)    & 0&77(3)      
   & 0&342(2)& 31(4)   &  1&35(6) & 1&36(12)  \\
\hline
 C & 0&6  &  0&18  & 0&3524(18)& 0&36(2)   & 1&36(10)     
   & 0&86(8) & 36(6)   &  2&4(3)  & \multicolumn{2}{c|}{}  \\
DD & 0&6  &  0&18  & 0&3389(13)& 0&339(16) & 1&31(7)      
   & 0&86(11)& 38(4)   &  2&5(3)  & 2&3(2.4)  \\
\hline
\end{tabular}
\end{center}
\end{table}
%

Tuning the scalar and fermion mass is, of course, important in
order to be close to the critical line separating the broken (FM) and
symmetric (PM) phases \cite{LIMOWI}, and at the same time avoid strong
finite size effect.
This prevents us from going on the $6^3 \cdot 12$ lattice to a very
small Yukawa coupling $G_\psi \ll 1$, because then the fermion
mass becomes too small.
At the strongest Yukawa coupling (at $\kappa=0$) the minimum
of the scalar mass on our lattice is quite large (above 1),
therefore we could not achieve the desired value
$m_{R\sigma} \simeq 0.6 - 0.8$.
This is similar to the behaviour observed at $\lambda = \infty$
\cite{FLMMPTW}.
More generally, in the investigated range of $G_\psi$ the
qualitative change of the renormalized quantities between
$4^3 \cdot 8$ and $6^3 \cdot 12$ lattices is quite similar
to the one at $\lambda=\infty$.
This allows to choose the points with label c,d and g as optimal in the
three sets of points for the lower bound in table \ref{t1}.
These are shown in figs.\ 1 and 2 together with the one-loop
perturbative estimates of the vacuum stability lower limit
at the given cut-offs.
In these figures also the data at $\lambda=\infty$ are included, which
show the behaviour of the triviality upper bound for the renormalized
quartic coupling.
For completeness we display the results for $\lambda = \infty$ in table
\ref{t1} too, because more statistics has been collected for some points
since publication of \cite{FLMMPTW}.

The agreement between numerical simulation data and the one-loop
perturbative estimates is remarkably good.
In particular, at the strongest couplings, within our errors, there
is practically no difference in the renormalized couplings between
$\lambda \simeq 0$ and $\lambda=\infty$.
This means that the strong Yukawa coupling alone is able to
induce the maximal possible renormalized quartic coupling at the
given cut-off.

Table \ref{t1} also includes results for the renormalized Yukawa
coupling $G_{R \psi}^{(3)}$, defined in terms of a three-point function
(see \cite{FLMMPTW}).
At tree level, and moreover in the one-loop approximation, it coincides
with $G_{R \psi}$.
The numerical results show that both couplings are consistent
with each other, even in those points where the errors are relatively
large.


\section{Yukawa coupling}                                 \label{s3}

One can see in table \ref{t1} that in the broken (FM) phase the
renormalized Yukawa coupling $G_{R\psi}$ grows with the bare coupling
$G_\psi$ roughly linearly, up to quite strong values above the tree
unitarity limit $ \sqrt{2\pi} \simeq 2.5$.
In previous numerical simulations in the symmetric (PM) phase of the
$\rm U(1)_L \otimes U(1)_R$ \cite{FKLMMM} and
$\rm SU(2)_L \otimes SU(2)_R$ \cite{LINWIT} symmetric Yukawa models
very strong renormalized Yukawa couplings were observed, as well.
We would like to see how strong the renormalized Yukawa couplings can
be in the PM phase in the mirror-fermion decoupling limit.
Since we do not want to go to the region with $\kappa < 0$ where
reflection positivity could not be proven \cite{LMMW},
the maximal possible value of $G_{R\psi}$
should come from tuning $G_\psi$ along the $\kappa=0$ axis in the PM
phase.
We tune $G_\psi$ such that the physical scalar mass $m_\phi$ is around
$0.7$ on the $6^3\cdot 12$ lattice to avoid large finite size effects.
We also run the same point on $8^3\cdot 12$ and $8^3\cdot 16$ lattices
to see how various quantities (especially the renormalized Yukawa
couplings) change with different volumes.

The renormalized Yukawa coupling matrix
$G_R\equiv{\rm diag}(G_{R\psi},G_{R\chi})$ is defined as
\be \label{eq03}
G_R\equiv -{m_R^2\over 4\sqrt{Z_R}}\, \tilde\Gamma_R(k)\,
Z_\Psi^{-1/2}\, \langle\Phi\Psi\bar{\Psi}\rangle_0\,Z_\Psi^{-1/2}\,
\tilde\Gamma_R(k) \,,
\ee
where the three-point Green's function
$\langle\Phi\Psi\bar\Psi\rangle_0$ and $Z_\Psi^{-1/2}$
are defined in \cite{LINWIT}.
$\tilde\Gamma_R(k)$ is the momentum-space renormalized fermionic
two-point vertex function defined at the smallest momentum
$k=(\vec 0,k_4)$, $k_4=\pi/T$.
Near $k=0$ the behaviour of $\tilde{\Gamma}_R(k)$ is
\be \label{eq04}
\tilde\Gamma_R(k)\,\, \simeq
\,\, i\gamma_4\, \bar{k}_4\,+\,M_R\,\,,
\hspace{2em}
\bar{k}_4\equiv {\rm sin} k_4={\rm sin}({\pi\over T})\,\,,
\hspace{2em}
M_R= \left( \begin{array}{cc}
   0  &  \mu_R  \\
   \mu_R  & 0
\end{array}  \right) \,,
\ee
where $\mu_R$ is the renormalized fermion mass.
In previous work we did the leading-order approximation by neglecting
the term $i\gamma_4\bar k_4$ in the measurement since $T$ is
large.
However, from data in \cite{LINWIT}, we suspect that it might
cause some small but visible effects on $G_R$.
Therefore we decided to improve the measurement of $G_R$ by including
this leading-order correction.
Also, we remove a previous inconsistency in normalization in the
definition of $G_R$ in the PM phase.
Before, the convention was such that
the full scalar propagator at zero momentum
$\tilde G(0)\equiv \sum_{x,y}\langle \phi_{Sx}\phi_{Sy}\rangle/L^3T$,
after renormalization, was normalized to $1/m_R^2$.
However, this does not correspond to
the natural normalization convention we used in the broken phase
\cite{FLMMPTW}.
There the convention corresponds to the one in the
PM phase such that $\tilde G(0)$ is renormalized to $4/m_R^2$.
In order to have a fair comparison of the renormalized Yukawa
couplings in both phases, we decide to switch to this new convention of
normalization for $G_R$ shown in (\ref{eq03}).
This implies that data on $G_R$'s in \cite{LINWIT} should be scaled
down by a factor of two in this new normalization convention.
(A similar inconsistency in the $\rm U(1)_L \otimes U(1)_R$ symmetric
model \cite{FKLMMM} can be removed by dividing the values of $G_R$ there
by a factor of $\sqrt{2}$.)

One should notice that, as shown in (\ref{eq03}), scalar quantities
appearing in the definition of $G_R$ should be the renormalized scalar
mass $m_R$ and the wave function renormalization factor $Z_R$
defined at vanishing momentum \cite{MONMUN}.
However, the scalar mass we actually use in the measurement
is the physical scalar mass $m_\phi$  obtained
from a cosh fit of the scalar field correlation function along the
time direction.
The wave function renormalization factor $Z_3$, which is defined
through the residuum of the pole of the propagator, cannot be determined
reliably from our statistics.
Therefore we define the wave function renormalization factor $Z_\phi$
in terms of the mass $m_\phi$ and the susceptibility $\tilde G(0)$ by
means of
\be \label{eq05}
Z_\phi = m_{\phi}^2 \, \tilde G(0) \,.
\ee
In general, $m_R$ and $Z_R$ is different from $m_\phi$ and $Z_\phi$.
In a weakly interacting system (e.g.\
pure scalar $\lambda\phi^4$ theory), they
are very close to each other\cite{MONMUN}.
In our SU(2) mirror-fermion model, the bare Yukawa coupling
$G_\psi$ is large, therefore there is no
guarantee that they are still close to each other.
On the other hand, the correct measurement of $m_R$ and $Z_R$ is
crucial to the measurement of $G_{R\psi}$ and $G_{R\chi}$, we therefore
also measure $m_R$ and $Z_R$
to see how they differ from $m_\phi$ and $Z_\phi$.
Since we are on the finite lattice, we estimate $Z_R$ and $m_R$ from
\be \label{eq06}
Z_R^{-1}\equiv [ \tilde G(k)^{-1}-\tilde G(0)^{-1} ]/\hat k^2\,\,,
\hspace{2em}
m_R^2={Z_R\over \tilde G(0)} \,,
\ee
where
\be \label{eq07}
k=(\vec 0,k_4)\,\,,
\hspace{2em}
k_4={2\pi\over T}\,\,,
\hspace{2em}
\hat k^2=4{\rm sin}^2({k_4\over 2}) \,.
\ee

The results of the numerical simulations in the symmetric phase are
collected in table \ref{t2}.
{}From the table one can clearly see that $G_{R\chi}$ is now
zero within error bars as expected from the shift symmetry at
$G_\chi=0$ \cite{GOLPET,LINWIT}.
This indicates that the term we used to neglect
does, indeed, have some visible effect on the renormalized Yukawa
couplings.
Meanwhile, the natural definition of the renormalized couplings are
taken at zero momentum since there is no infrared singularity in
the PM phase.
But $G_R$'s we measured according to (\ref{eq03}) are actually
defined at $k=(0,0,0,\pi/T)$, which will approach
zero in the infinite $T$ limit.
If the inverse propagator had a non-negligible curvature near zero
momentum, this would influence the determination of $Z_R$ and hence of
other renormalized quantities.
We therefore took measurements at two different values of $T$,
i.e.: $T=12, 16$, to see how we can get $G_R$'s at zero
momentum by extrapolation.
Our data show that $G_{R\psi}$ and $G_{R\chi}$ basically stay
unchanged as $T$ goes from 12 to 16.
We therefore believe that curvature effects in the propagator are not
large and that the data we have on $G_R$'s are, to a good
approximation, the renormalized Yukawa couplings defined at zero
momentum.

The fluctuations of $Z_R$ and $m_R$ are quite large, as expected.
Within error bars, they agree with $Z_\phi$ and $m_\phi$.
This shows that our measurement of $G_{R\psi}$ and
$G_{R\chi}$ using $m_\phi$ and $Z_\phi$ can be taken as a good
approximation to the $G_R$'s defined in (\ref{eq03}).

As one can see from table \ref{t2}, on our lattices the maximal value
of $G_{R\psi}$ in the region of non-negative scalar hopping parameter
is more than twice the value of the tree unitarity bound.
This supports our previous results in the symmetric phase
on smaller lattices and less statistics \cite{FKLMMM,LINWIT}.
With the present correct normalization the difference between the
maximal value of $G_{R\psi}$ in the symmetric and broken phase
is not dramatic, although the values are still larger in the
symmetric phase (see tables \ref{t1}--\ref{t2}).

%
\begin{table}[tb]
%
%
\caption{  \label{t2}
The main renormalized quantities at $\lambda=\infty$, $G_\chi=0$,
$G_\psi=1.09$, $\kappa=0$ and $K=2/19$ in the PM phase on
various lattices.
Data on $6^3\cdot 12$ lattices are collected from 20000 trajectories
while on $8^3\cdot 12$ and $8^3\cdot 16$ we have about 6000
trajectories each.}
\begin{center}
\begin{tabular}
{|c|r@{.}l|r@{.}l|r@{.}l|r@{.}l|r@{.}l|r@{.}l|r@{.}l|}
\hline
Size&\multicolumn{2}{c|}{$m_\phi$} & \multicolumn{2}{c|}{$m_R$}
&\multicolumn{2}{c|}{$Z_\phi$} & \multicolumn{2}{c|}{$Z_R$}
&\multicolumn{2}{c|}{$\mu_R$}  & \multicolumn{2}{c|}{$G_{R\psi}$}
&\multicolumn{2}{c|}{$G_{R\chi}$} \\
\hline
$6^3\cdot 12$
   & 0&69(3) &  0&60(17) & 1&52(5) & 1&11(38)     & 0&664(5)
   & 5&84(16)&   -0&03(7) \\
$8^3\cdot 12$
   & 0&65(7) &  0&56(11) & 1&84(15)& 1&36(57)     & 0&654(9)
   & 6&04(37)&   -0&12(17) \\
$8^3\cdot 16$
   & 0&55(3) &  0&51(11) & 1&71(33)& 1&44(73)     & 0&63(1)
   & 5&87(54) &   0&07(9) \\
\hline
\end{tabular}
\end{center}
\end{table}
%

\section{$1/N$ expansion}                                 \label{s4}

The results of numerical studies of the phase diagram of the model
at both $\lambda=\infty$ and $\lambda=10^{-6}$ are reported in
\cite{FLMMPTW}.
The particular interest in the phase structure at small $\lambda$ arose
after it was reported \cite{AJS} that models with naive fermions do not
exhibit a ferrimagnetic (FI) phase and that instead a first order phase
transition is observed.
Clearly the absence of any such transition is of great importance for
the study of bounds on the couplings within our model.

Our Monte Carlo investigations show that at least the physically
relevant phase transition from the PM to the FM phase is second order.
In particular, the magnetization varies smoothly across the transition.

The numerical analysis of the phase diagram was supplemented by a
$1/N_f$ expansion in leading order, where $N_f$ denotes the number
of fermion--mirror-fermion doublet pairs.

Our strategy was to calculate the one-loop effective potential to
leading order in $1/N_f$ as a function of the fluctuation field
\be \label{eq08}
  \sigma_x \equiv \phi_{4x}-s-(-1)^{x_1+x_2+x_3+x_3}\,\widehat{s} \,,
\ee
where $\phi_{4x}$ is the fourth real component of the scalar field
$\varphi_x$, and $s$ and $\widehat{s}$ are the positions of the minimum
of the effective potential at tree level with respect to $\varphi_x$
and the staggered scalar field $\widehat{\varphi}_x$, respectively.
After performing a Fourier transformation to momentum
space, we finally obtain the effective potential
$V_{\rm eff}[\widetilde{\sigma}(0),\,\widetilde{\sigma}(\pi)]$ as
a function of the fluctuation field at both the zero- and
$4\pi$-corners of the Brillouin zone.
The qualitative features of the
transitions from PM to FM (PM to AFM \cite{LIMOWI}) can now be
studied via the  dependence of $V_{\rm eff}$ on the field
$\widetilde{\sigma}(0)$ (the field $\widetilde{\sigma}(\pi)$).

A first observation is that to leading order in $1/N_f$ the effective
potential is a quadratic function of $\widetilde{\sigma}(0)$ and
$\widetilde{\sigma}(\pi)$, and therefore we do not expect a first
order phase transition which would rather require a quartic term,
resulting in a double-well structure of the potential.

For $\lambda=0$ we obtain estimates for the critical value of the
scalar hopping parameter $\kappa$ from the two gap equations
\begin{eqnarray} \label{eq09}
-16\,\kappa_{\rm cr} + 2 - 8N_f\,G_\psi^2 \int_q\,
\frac{\overline{q}^2}{(\overline{q}^2+\mu_q^2)^2+G_\psi^2\,s^2\,
                       \overline{q}^2} & = & 0
                 \qquad ({\rm for\ PM \leftrightarrow FM}),
 \\ \label{eq10}
               16\,\kappa_{\rm cr} + 2 & = & 0
                 \qquad ({\rm for\ PM \leftrightarrow AFM}),
\end{eqnarray}
where $\overline{q}_\mu=\sin(q_\mu)$, and the integral is taken over
the Brillouin zone.
It is therefore only the transition PM to FM which
is affected by fermionic contributions in leading order.
Its transition line bends down for increasing Yukawa couplings
and finally intersects the straight transition line from PM to AFM at
$\kappa_{\rm cr} = -1/8$.
Obviously, in leading order of the $1/N_f$ expansion
the qualitative behaviour of the phase transition lines is
similar to all our numerical studies \cite{FLMMPTW,LIMOWI}.

According to the $1/N_f$ expansion the FI phase exists in
leading order, since solutions to the minimum of $V_{\rm eff}$ at
non-zero values of both $\widetilde{\sigma}(0)$ and
$\widetilde{\sigma}(\pi)$ are found to exist.
Furthermore, the position of the minimum of $V_{\rm eff}$ varies
smoothly across all phase transition lines, therefore suggesting the
existence of second order phase transitions only.

The findings from the $1/N_f$ expansion together with the Monte Carlo
results are strong evidence that the allowed region of renormalized
couplings can safely be studied at $\lambda\simeq0$ as well.


\section{Conclusions}                                     \label{s5}

The main conclusion of the numerical simulations of heavy fermions
in the $\rm SU(2)_L \otimes SU(2)_R$ symmetric Yukawa model
is that the estimates of the upper and lower bounds on renormalized
couplings obtained in one-loop perturbation theory work well.
All qualitative features of the one-loop $\beta$-functions are
supported, including the fixed point in the ratio of the fermion
to scalar mass $\mu_{R\psi}/m_{R\sigma}$ (see fig.\ 2).
The upper limit of the renormalized couplings (at the fixed
ratio of $g_R/G_{R\psi}^2$) is provided by our requirement of
reflection positivity.
If this would not be imposed, the line towards the upper right
corner of fig.\ 1 would be continued, but probably not much
beyond our values.
The reason is the appearance of the ferrimagnetic (FI) phase
transition in the region of negative scalar hopping parameters,
somewhat beyond the maximal bare Yukawa coupling we consider.
The existence of the FI phase at small and large values of the
bare quartic coupling $\lambda$ is implied by both numerical
simulations and $1/N$ expansion.
These conclusions are also supported in a recent numerical
simulation of the same continuum ``target theory'' as ours, by
using reduced staggered fermions \cite{BOFRSMVI}.

Although the observed qualitative behaviour is certainly consistent
with the one-loop perturbative scenario implying the triviality
of the continuum limit, one has to keep in mind that present
simulations are done at relatively low cut-offs.
In particular, the evolution of the couplings towards smaller
values at decreasing cut-offs should be investigated in the
future.
At present the renormalized Yukawa coupling can have quite large
values (see tables \ref{t1}--\ref{t2}).
In the symmetric phase it reaches more than twice the tree level
unitarity bound with both scalar and fermion masses about equal
to 0.5 in lattice units.


\vspace{1cm}
\large\bf Acknowledgements \normalsize\rm \newline
\vspace{3pt}

\noindent
We thank Christoph Frick for useful discussions.
The Monte Carlo calculations for this work have been performed
on the CRAY Y-MP/832 of HLRZ J\"ulich.

\newpage

\newpage
%
\begin{center}     \Large\bf Figure captions \normalsize\rm
\end{center}

\vspace{15pt}
\bf Fig.\,1.    \hspace{5pt} \rm
The data for the upper and lower bounds on $g_R$ as a function of
$G_{R\psi}^2$ together with the perturbative estimates for scale
ratios $\Lambda/m_{R\sigma}=3$ (solid curve) and $\Lambda/m_{R\sigma}
=4$ (dotted curve).
Open points denote the data for the lower bound (points c, d, g in table
\ref{t1}), whereas full symbols are data for the upper bound.
The $6^3\cdot12$-lattice is represented by triangles, whereas points on
$8^3\cdot16$ are denoted by squares.

\vspace{15pt}
\bf Fig.\,2.    \hspace{5pt} \rm
The mass ratio $\mu_{R\psi}/m_{R\sigma}$ as a function of $G_{R\psi}$
in comparison with one-loop perturbative estimates for
$m_{R\sigma} = 0.75$ (dotted curve), $m_{R\sigma}=1$ (full curve) and
$m_{R\sigma} = 1.25$ (dashed curve).
The solid horizontal line represents the fixed point at infinite
cut-off.
The explanation of symbols is as in Fig.\ 1.

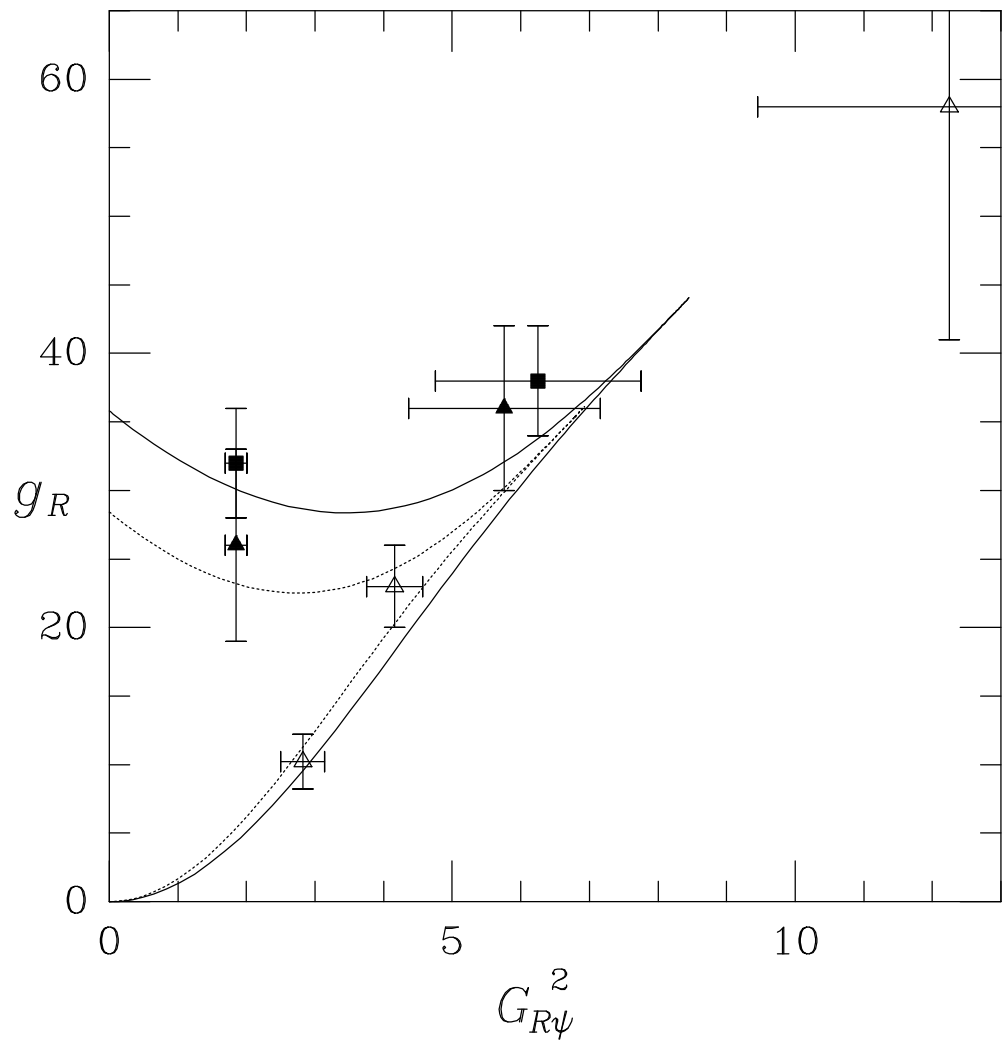
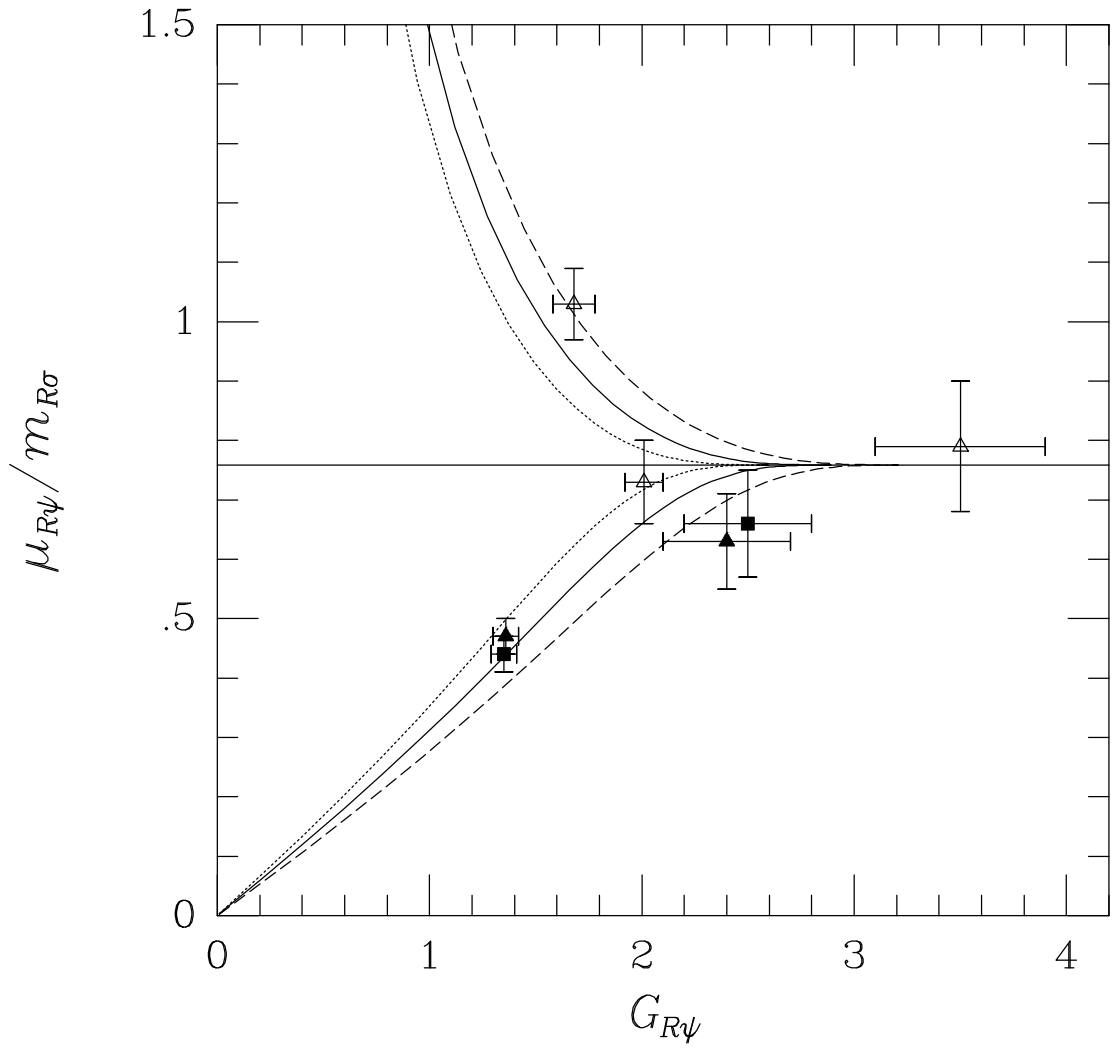
\end{document}